\documentclass{aipproc}

\usepackage{epsfig}

\layoutstyle{6x9}

\begin{document}

\title[Observability of EGRET tentative IDs by the next-generation CTs]{ 
        An observability 
	study for the tentatively identified 3EG sources likely 
        to be detected by the next-generation Cherenkov telescopes}

\author{Dirk Petry} {
   address={Dept. of Physics and Astronomy, Iowa State University, Ames, IA 50011, USA},
   email={petry@iastate.edu}
}

\author{Olaf Reimer}{
   address={Laboratory for High Energy Astrophysics, NASA/GSFC, Greenbelt, MD 20771, USA},
   email={olr@egret.gsfc.nasa.gov}
}

\copyrightyear{2001}

\begin{abstract}

 We present a compilation of data on the 22 tentatively identified gamma-ray 
sources from the Third EGRET Catalog
 which may be detected by the next-generation imaging atmospheric
 Cherenkov telescopes.

\end{abstract}

\date{}

\maketitle

\section{Introduction}

The Third EGRET Catalog (3EG, Hartman et al. 1999),
comprises 271 objects. Among these, 197 are not identified with a counterpart at
lower wavelengths (radio, optical or X-rays). Seven of these are now believed to be artefacts
of the background model near bright sources. The remaining 190 are the unidentified
EGRET objects (UNIDs). A sizable number of researchers is working on identifying the UNIDs
and so far more than 38 have a published tentative ID which still needs to be confirmed
by either more observations or improved analysis of archival data. We call these sources 
the tentatively identified EGRET objects (TIDs).

New contributors to the field will be the
next-generation Cherenkov telescope (CT) observatories which are under construction
in Australia (CANGAROO III, e.g. Mori et al. 1999), Namibia (HESS I, e.g. Hofmann et al. 1999),
La Palma (MAGIC I, e.g.  Lorenz et al. 1999) and Arizona (VERITAS, e.g. Krennrich et al. 1999).

These new instruments will reach thresholds  below 100 GeV and source location accuracies
of about 1'. All UNIDs are unidentified because their position is only known with insufficient
accuracy, some of the position probability maps having 95\% confidence level contour radii
of more than 1$^\circ$. With an order of magnitude increase in location accuracy, deep well-targeted
observations in the radio, optical and X-ray range become possible and make an identification almost certain. 
In addition, the much improved photon statistics of CTs
(collection areas $>$ 10$^4$~m$^2$) result in a higher sensitivity for pulsed components and
thus Pulsar identifications.
However, CTs can only contribute for those sources
which show emission above several 10 GeV.

In Petry (2001), a catalog was compiled which contains all UNIDs which may possibly be detectable
by the next-generation Cherenkov telescopes under moderate assumptions about spectral steepening
and taking into account the elevation-dependent sensitivity of the instruments. 
This catalog contains 78 objects. Among them are 22 TIDs. These objects justify a closer examination
since for their tentative counterparts various pieces of information exist which are
not available for the other UNIDs:
 We have an exact source position which can be targeted.
 We know the source type and have therefore at least vague
      model predictions for the spectrum beyond the EGRET energy range.
 We have also model predictions for the variability characteristics
      of the source.

In this article we present first results of our data compilation and studies
concerning  the 22 TIDs
which may exhibit significant emission beyond 10 GeV and for which the 
next-generation Cherenkov telescopes may provide the
clue to their final identification.

\section{The data presented here}

Table 1 gives a summary of the data presented at the conference.
For each object we examine:
\begin{itemize}
\item What is the predicted emission of the TID in
   the energy regime near the threshold of the next-generation
   Cherenkov telescopes (CTs)?
\item Which of the four observatories can observe the object?
\item Is the emission variable?
\item What is the angular size of the tentative counterpart?
\item Are bright stars nearby which  may influence the sensitivity
   of the CTs? 
\item Are there neighbouring EGRET objects which may lead to source confusion?
\item What chances are there for a detection if the tentative identification
   turns out to be wrong? Will new pointings be necessary?
\end{itemize}
Due to the limited space in these proceedings, we refer for more detailed 
information to our poster which can be found in {\footnotesize  petry-reimer\_poster.eps.gz }
at {\footnotesize http://cossc.gsfc.nasa.gov/meetings/Gamma2001/session17/ } 
and viewed using ghostview, and to Petry \& Reimer (2001). 
We give here, however, the complete list of references.

\begin{figure}[hbt]
\centering
\epsfysize=7.4cm
\epsffile{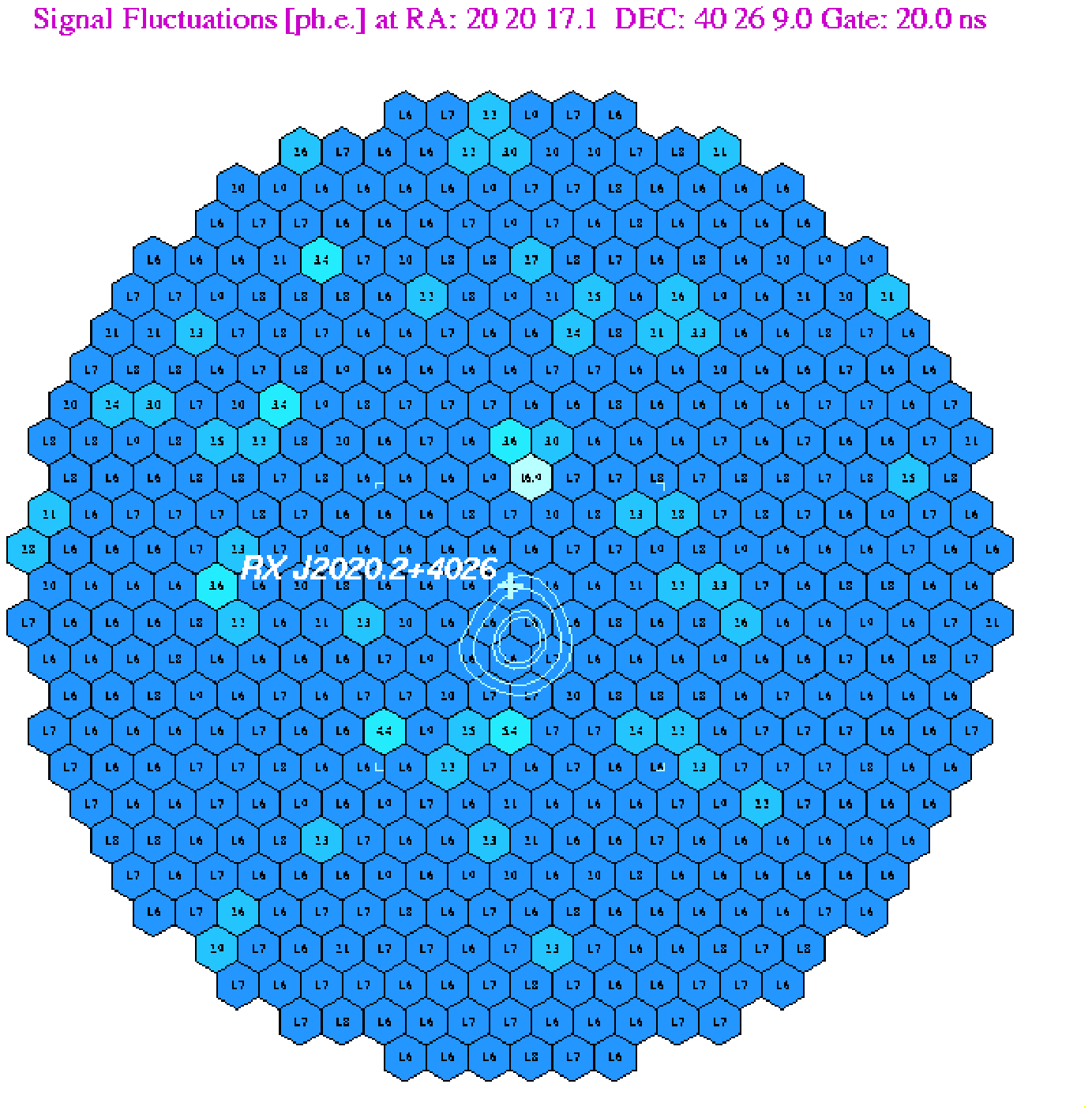}
\epsfysize=7.4cm
\epsffile{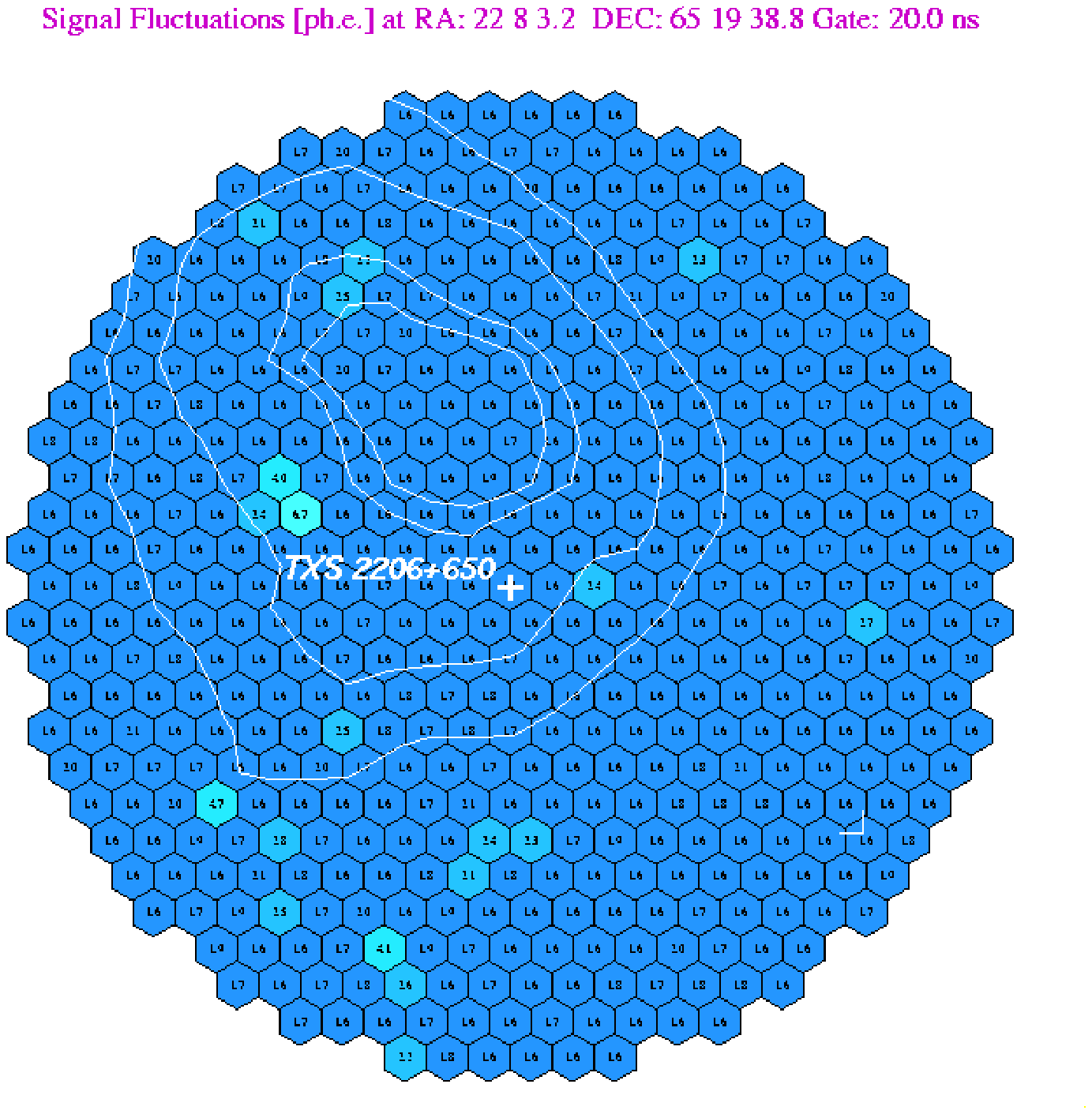}
\caption{Examples of EGRET position probability maps for two of the tentatively
identified sources discussed here: 3EG J2020+4017 (left) and 3EG J2206+6602
(right). The maps are superimposed on simulated representations of the field of view
of the VERITAS Cherenkov telescope camera. The numbers in the individual photomultiplier
pixels represent the expected signal fluctuations due to night sky background and star light 
in units of photoelectrons.}
\end{figure}

\section{Simulated Camera Responses}

We have examined the predicted response of the VERITAS photomultiplier
camera to the starfield at each of the 22 source positions.
The results were obtained from a simulation  of the VERITAS optics and 
wavelength-dependent photomultiplier response and are shown  here
for the first time.   The starfield information (star positions and spectra)
was extracted from the SKY2000 master star catalog (Sande et al. 1998) which is reasonably
complete up to magnitude 9. If not available from the catalog, the U band
magnitude was calculated from the B and V magnitude assuming a main sequence star.
Most important result of the simulation are the maps of the Poisson signal 
fluctuations in units of photoelectrons
caused by the starlight and diffuse NSB in the field of view around each source.
In order to not exceed the page limit of this
publication, we only show two examples (figure 1), one for a very extended
and one for a well constrained EGRET position probability map.

\begin{table}[!t]
{\footnotesize
\begin{tabular}{llccccccccc}
\hline
3EG name & tent. ID$^{(1)}$ & RA/DEC J2000$^{(2)}$     & l/b$^{(3)}$   & z$^{(4)}$ & $\alpha^{(5)}$ & $F$(60 GeV)$^{(6)}$ &   observ.$^{(7)}$ & var.$^{(8)}$ & s.$^{(9)}$  & p.$^{(10)}$ \\
\hline
 0010+7309 & RX J0007.0+7302  & 00 07 02.2 +73 02 59  & 119.66 +10.46 &   -               & -1.6$^b$    &  8.3E-16$^f$     &   M, V         & 0.31                  & no              & no \\
	   & in SNR CTA1 \\
 0222+4253 & 3C 66A & 02 22 39.6 +43 02 07.8 & 140.14 -16.77            & 0.444       & -2.0       & 3.1E-10     & M, V           &   $^e$              & no   & no \\
 0241+6103 & LSI+61$^\circ$303 &02 40 31.67 +61 13 45.6  & 135.68 +1.09     & -           & -2.2       & 1.3E-10     & M, V            & 0.49              & no  & no \\
 0617+2238 & 1WGA J0617.1+2221  & 06 17 06.1  +22 21 30   &  189.23 +2.90   & -           & -1.8$^b$   & 2.2E-10     & H, M, V         & 0.26              & yes & no \\
	  & in SNR IC443 \\
 0808+4844$^a$ & QSO B0809+483 & 08 13 36.09 +48 13 02.5 & 171.17 +33.24 & 0.87           & -2.3      & 2.9E-11     & M, V            & n/a             & no  & yes \\

 0812-0646$^a$ & PKS 0805-077 & 08 08 15.54 -07 51 09.9  & 229.04 +13.16 & 1.84           & -2.4      & 2.0E-11     & C, H, M, V      & n/a             & no   & yes \\
 0917+4427 & QSO B0917+449 & 09 20 58.46 +44 41 54.0 & 175.70 +44.82 &     2.18           & -2.1      & 2.8E-11     & M, V            &  n/a            & no  & yes \\
 1009+4855$^a$ & QSO B1011+496 & 10 15 04.23 +49 26 00.7 & 165.53 +52.71 & 0.20 &          -2.0$^d$ &    7.5E-11    & M, V            & n/a          & no   & yes \\
 1323+2200 & QSO B1324+224 & 13 27 00.86 +22 10 50.2  & 3.38 +80.53     & 1.40  &           -1.9   &   3.0E-10      & M, V            &  2.69         & no  & no \\
 1410-6147 & RX J1420.1-6049  & 14 20 07.8 -60 48 56  &   314.45 +1.38    &    -   &         -2.0$^b$ &   1.8E-10      & C, H            & 0.33        & yes  & no \\
	& in SNR 312.4-00.4 \\
 1800-2338 & SNR W28 & 18 01.0 -23 11                 & 6.71 -0.05 &          -    &        -2.0$^c$ &  2.2E-10        & C, H, M, V     & 0.03   & yes   & no\\
 1824-1514 & LS5039  & 18 26 14.9 -14 50 51   & 16.88 -1.29 &                    -     & -2.4           & 7.5E-11     &  C, H, M, V &       n/a  & yes   & yes \\
 1835+5918 & RX J1836.2+5925 & 18 36 13.82 +59 25 28.9  & 88.88 +25.00  &      -     & -1.7$^b$ &  $\approx$1.4E-15$^f$      & M, V &  0.15 &  no  & no \\
 1856+0114 & PSR 1853+01  &18 56 10.89 +01 13 20.6  & 34.56 -0.50 &          -       & -1.8 &        8.3E-15$^f$             & C, H, M, V & 0.80 & yes  & no \\
           & in SNR W44 \\
 1903+0550 & SNR 040.5-00.5 & 19 06.9 +06 33 & 40.52 -0.44&               -          & -2.5 &       3.7E-11          &   C, H, M, V     &  0.35 & yes & yes\\
 2016+3657 & TXS 2013+370  & 20 15 28.89 +37 10 58.7 & 74.87 +1.22 & ?                & -2.0$^c$ &            1.3E-10   & M, V             & 0.37  & yes! & no \\
 2020+4017 & RX J2020.2+4026  & 20 20 17.1 +40 26 09   & 78.09 +2.27 &    -          & -2.0$^b$ &       1E-17$^f$             & M, V           &  0.07  & yes!  & no \\
	   & in SNR G78.2 \\
 2100+6012$^a$ &  B2101+6003 & 21 02 40.31 +60 15 09.8   & 97.96 +9.01 &       ? & -2.1$^d$ &   3.5E-11                & M, V           & 0.15  & yes  & no\\
 2206+6602$^a$ & TXS 2206+650 & 22 08 03.20 +65 19 38.8 & 106.94 +7.66 & ?        & -2.3 &          2.6E-11             & M, V           & 0.27  & yes  &  yes \\
 2227+6122 & RX J2229.0+6114 & 22 29 04.97 +61 14 12.9 & 106.65 +2.95 &     -      & -2.2 &         6.1E-11             & M, V          & 0.10  & no  & no \\
 2255+1943$^a$ & QSO B2246+2051 & 22 53 07.36 +19 42 34.8 & 88.33 -35.09 & 0.284  & -2.3$^d$ &      4.3E-11             & C, H, M, V    &  0.41  & no  & yes \\
 2352+3752$^a$ & QSO 2346+385 & 23 49 20.91 +38 49 17.6 & 109.89 -22.47 & 1.03    & -2.6$^d$ &      1.3E-11             & M, V          & 24.92  & no  & yes \\
\end{tabular}
\caption{\label{tab-thelist} \footnotesize The tentatively identified EGRET sources likely to be detected by the
 next-generation Cherenkov telescopes $[$Column titles: (1) name of the tentative counterpart, 
(2) equatorial coordinates of the counterpart, (3) galactic coordinates of the counterpart,
(4) redshift for the extragalactic counterparts if known, (5) spectral index at 100 MeV from 3EG, 
(6) expected integral flux above 60 GeV  (cm$^{-2}$s$^{-1}$),
(7) source observability (C = CANGAROO, H = HESS, M = MAGIC, V = VERITAS), (8) variability index 
of the 3EG source as defined in Tompkins (1999), 
(9) problematic starfield?, (10) new pointings be
necessary if the tentative identification turns out to be wrong?$]$
$[$Abbreviations: $^a$ will probably require $>$ 50 h of observation time (Petry 2001), 
$^b$ high energy cutoff visible in EGRET data, $^c$ soft low energy tail , $^d$ low statistics in EGRET data, 
$^e$ short-time variability observed, $^f$ predicted differential flux at 60 GeV (cm$^{-2}$s$^{-1}$MeV$^{-1}$), for comparison: predicted
Crab Nebula flux at 60 GeV is $1.28 \times 10^{-14}$cm$^{-2}$s$^{-1}$MeV$^{-1}$.$]$ 
}
}
\end{table}

\section{Conclusion}
\vspace{-0.2cm}
The next-generation Cherenkov telescopes (CTs)
will be able to make an important contribution to the identification of some of the
enigmatic unidentified sources of the third EGRET catalog. EGRET UNIDs for which
a tentative identification exists are especially easy to target, and instruments
on the northern hemisphere will be able to observe almost all such sources for which
emission beyond 30 GeV can be expected. The short catalog of 22 such sources which
we have compiled here, shows that a positive detection of any of these objects
by CTs will be an interesting result in itself providing constraints 
for source models and, of course, leading to a clear identification of the 3EG source.
Furthermore, the lessons learned form the observations of these objects will help
in the examination of the remaining 57 EGRET UNIDs from the list compiled in
Petry (2001)  which have no identification whatsoever but which may have
significant emission beyond 30 GeV. 

For the beginning of an observation campaign, the most interesting object 
in our list is 3EG J1856+0114.
This object has a flat spectrum with no obvious cut-off below 10 GeV.
Due to its proximity to the SNR W44, it has been studied extensively (see e.g. the overview in
Buckley et al. 1998). W44 is a radio shell-type SNR with an angular diameter of about 0.5$^\circ$ 
associated with PSR 1853+01. There is both evidence for a synchrotron nebula and interactions
with molecular clouds. Extrapolating the thin outer gap model of Zhang \& Cheng (1998) to
60 GeV yields a differential flux of 65~\% of the Crab Nebula.
The object is an ideal candidate for CT observations and is accidentally the only UNID 
which can be equally well observed both from the southern and the northern hemisphere. 
It could therefore be used to cross-calibrate the four CT observatories.

\begin{theacknowledgments}
\vspace{-0.2cm}
{\footnotesize D.P. would like to thank C. Duke, Grinnell College, Iowa, for contributing ray tracing code
for the VERITAS optics simulation.
This research has made use of the SIMBAD data\-base, operated at CDS, Strasbourg, France. }

\vspace{-0.8cm}

\end{theacknowledgments}


\begin{thebibliography}{1}
\bibitem{a}
Bocchino, F. \& Bykov, A.M., 2000,  A\&{}A, 362, L19


\bibitem{b}
Bloom, S.D., et al., 1997, ApJ, 488, L23


\bibitem{c}
Brazier, K.T.S., et al., 1996, MNRAS, 281, 1033 


\bibitem{d}
Brazier, K.T.S., et al., 1998, MNRAS, 295, 819


\bibitem{e}
Buckley, J.H., et al., 1998, A\&{}A, 329, 639


\bibitem{f}
Cheng, K.S. \& Zhang, L., 1998, ApJ, 498, 327


\bibitem{g}
Fichtel, C.E., et al., 1994, ApJS, 94, 551


\bibitem{h}
Halpern, J.P., et al., 2001, ApJ, 547, 323


\bibitem{i}
Halpern, J.P., et al., 2001, ApJ, 551, 1016


\bibitem{j}
Hartman, R.C., et al., 1999, ApJS, 123, 79


\bibitem{k}
Hofmann, W., et al., 1999, in Dingus, B.L., et al. (eds.) Proc. 
``Towards a Major Atmospheric Cherenkov Detector VI'', AIP Proc.  515, 500

 
\bibitem{l}
Keohane, J.W., et al., 1997, ApJ, 484, 350


\bibitem{m}
Kniffen, D.A., et al., 1997, ApJ, 486, 126 


\bibitem{n}
Krennrich, F., et al., 1999, in Dingus, B.L., et al. (eds.) Proc. 
``Towards a Major Atmospheric Cherenkov Detector VI'', AIP Proc., 515, 515


\bibitem{o}
Kuiper, L. et al., 2000, A\&{}A, 359, 615 

\bibitem{p}
Lorenz, E., et al., 1999, in Dingus, B.L., et al. (eds.) Proc. 
``Towards a Major Atmospheric Cherenkov Detector VI'', AIP Proc., 515, 510


\bibitem{q}
Mattox, J.R.,  et al., 1997, ApJ, 481, 95


\bibitem{r}
Mattox, J.R., Hartman, R.C. \& Reimer, O., 2001,  ApJS, 135


\bibitem{s}
Merck, M., et al., 1996, A\&{}ASS, 120, 465


\bibitem{t}
Mirabal, N., et al., 2000, ApJ, 541, 180


\bibitem{u}
Mirabal, N., et al., 2001, ApJL, 547, 137


\bibitem{v}
Mori, M., et al., 1999, in Dingus, B.L., et al. (eds.) Proc. 
``Towards a Major Atmospheric Cherenkov Detector VI'', AIP Proc.,  515, 485


\bibitem{w}
Mukherjee, R., et al., 1997, ApJ, 490, 116


\bibitem{x}
Mukherjee, R., et al., 2000, ApJ, 542, 740 


\bibitem{y}
Neshpor, Yu. I., et al.,  2000, Astronomy Report, 44, 641


\bibitem{z}
Olbert, C.M., et al., 2001,  ApJL, in press (astro-ph/0103268)  


\bibitem{aa}
Paredes, J.M., et al., 2000, Science, 288, 2340

\bibitem{bb}
Petry, D., 2001, in Carrami\~{n}ana, A., Reimer, O. \& Thompson, D. (eds.), ``The nature of 
unidentified high-energy $\gamma$-ray sources'', IAU colloquia proceedings, in press (astro-ph/0101496)

\bibitem{bb1}
Petry, D. \& Reimer, O., 2001, Astropart. Phys., in preparation

\bibitem{cc}
Reimer, O., et al., 2000, Proc. 5th Compton Symp., AIP 510


\bibitem{dd}
Reimer, O., et al., 2001, MNRAS, 324, 772


\bibitem{ee}
Roberts, M.S.E., et al., 2000, in Carrami\~{n}ana, A., Reimer, O. \& Thompson, D. (eds.), ``The nature of 
unidentified high-energy $\gamma$-ray sources'', IAU colloquia proceedings, in press (astro-ph/0102471)


\bibitem{ff}
Romero, G.E., et al., 1999, A\&{}A, 348, 868


\bibitem{gg}
Rowell, G.P., et al., 2000, A\&{}A, 359, 337


\bibitem{hh}
Sande, C.B., et al., 1998, ``SKY2000 - Master Star Catalog - Star Catalog Database, Version 2'',
   Goddard Space Flight Center, Flight Dynamics Division 


\bibitem{ii}
Strickman, M.S., et al., 1998, ApJ, 497, 419


\bibitem{jj}
Tavani, M. et al., 1997,  ApJ, 497, L89 


\bibitem{kk}
Tompkins, W., 1999, PhD Thesis, Stanford University


\bibitem{ll}
Yadigaroglu, I.A. \& Romani, R.W., 1997, ApJ, 476, 347


\bibitem{mm}
Zhang, L. \& Cheng, K.S., 1998, A\&A, 335, 234


\bibitem{nn}
Zhang, L., Zhang, Y.J. \& Cheng, K.S., 2000, A\&{}A, 357, 957

\end{thebibliography}
\end{document}